# Advancements in Image Resolution: Super-Resolution Algorithm for Enhanced EOS-06 OCM-3 Data

Ankur Garg*, Tushar Shukla, Purvee Joshi, Debojyoti Ganguly,
Ashwin Gujarati, Meenakshi Sarkar, Dr. KN Babu, Dr. Mehul Pandya,
Dr. S. Manthira Moorthi, Debajyoti Dhar

agarg@sac.isro.gov.in
Space Applications Centre, Ahmedabad

**Abstract.** The Ocean Color Monitor-3 (OCM-3) sensor is instrumental in Earth observation, achieving a critical balance between high-resolution imaging and broad coverage. This paper explores innovative imaging methods employed in OCM-3 and the transformative potential of super-resolution techniques to enhance image quality. The super-resolution model for OCM-3 (SOCM-3) addresses the challenges of contemporary satellite imaging by effectively navigating the trade-off between image clarity and swath width. With resolutions below 240 meters in Local Area Coverage (LAC) mode and below 750 meters in Global Area Coverage (GAC) mode, coupled with a wide 1550-kilometer swath and a 2-day revisit time, SOCM-3 emerges as a leading asset in remote sensing. The paper details the intricate interplay of atmospheric, motion, optical, and detector effects that impact image quality, emphasizing the necessity for advanced computational techniques and sophisticated algorithms for effective image reconstruction. Evaluation methods are thoroughly discussed, incorporating visual assessments using the Blind/Referenceless Image Spatial Quality Evaluator (BRISQUE) metric and computational metrics such as Line Spread Function (LSF), Full Width at Half Maximum (FWHM), and Super-Resolution (SR) ratio. Additionally, statistical analyses, including power spectrum evaluations and target-wise spectral signatures, are employed to gauge the efficacy of super-resolution techniques. By enhancing both spatial resolution and revisit frequency, this study highlights significant advancements in remote sensing capabilities, providing valuable insights for applications across cryospheric, vegetation, oceanic, coastal, and domains. Ultimately, the findings underscore the potential of SOCM-3 to contribute meaningfully to our understanding of fine-scale oceanic phenomena and environmental monitoring.

**Keywords:** Superresolution, OCM-3, LAC, GAC, MTF, Deep Learning.

## 1    Introduction

In satellite remote sensing, a fundamental trade-off exists between image resolution and swath width, as shown in Table 1, which compares various contemporary sensors.

For instance, MERIS achieves a high resolution of 300 meters in LAC mode but offers a narrower swath of 1150 kilometers with a 3-day revisit time. Conversely, MODIS provides a wide swath of 2330 kilometers and a revisit time of 1-2 days, albeit with varying resolutions. VIIRS prioritizes a 3040-kilometer swath at 375 meters resolution and a quick 12-hour revisit time.

**Table 1.** Comparison of Resolution and Swath among Contemporary Satellites.

| Sensor | Resolution LAC (m) | Resolution GAC (m) | Swath (km) | Temporal Revisit |
|--------|--------------------|--------------------|------------|------------------|
| MERIS | 300 | 1200 | 1150 | 3 days |
| MODIS | 250(B1-B2),500(B3-B7), 1000(B8-B36) | - | 2330 | 1-2 days |
| VIIRS | 375(B I-IV), 750(M-15) | - | 3040 | 12 hours |
| OLCI | 300 | 1000 | 1270 | 2 days |
| OCM-3 | 360 | 1080 | 1550 | 2 days |
| SOCM-3 | <240 | <750 | 1550 | 2 days |

OLCI and OCM-3 maintain a balance, with OLCI offering 300 meters resolution and a 1270-kilometer swath for a 2-day revisit, while OCM-3 features 360 meters resolution and a 1550-kilometer swath, also with a 2-day revisit. SOCM-3 distinguishes itself by delivering superior resolution below 240 meters in LAC mode, below 750 meters in GAC mode, and a 1550-kilometer swath with a 2-day revisit time, marking a significant advancement in both spatial detail and observational frequency.

A typical image formation model is given in Figure 1. In the realm of super-resolution modeling, the focus is on the intricate task of reconstructing a high-resolution image $I'^{SR}$, from a given set of multiple low-resolution images $I_k^{LR}$. The complexity arises from the fact that each low-resolution image in the set has undergone a series of degradations throughout the imaging process, contributing to the overall loss of spatial details and image quality.

The super-resolution model focuses on reconstructing a high-resolution image from multiple low-resolution inputs, addressing various degradations like atmospheric distortions, motion effects, optical imperfections, downsampling, and photon-induced noise. By leveraging these degradation factors, the model employs advanced algorithms to reverse the effects and generate a high-quality image that accurately represents the original scene.

Super resolution enhances image resolution beyond its original capabilities, requiring several key elements. First, the input low-resolution images must be undersampled, providing fewer pixels than a higher-resolution version to allow for improvement. Second, variation in phase, or sub-pixel shifts in image features, is essential for extracting additional details, enabling the algorithm to infer finer nuances.

Mathematically, the ideal super resolution factor "s" represents the ratio of the desired resolution to the original; the number of samples needed is proportional to $s^2$, while required sub-pixel shifts are inversely related d=1/s. Additionally, the opti-cal system's Modulation Transfer Function (MTF) must maintain non-zero values up to the Nyquist frequency of the desired resolution, ensuring that fine details are captured. Collectively, these factors enhance the effectiveness of super resolution techniques in improving image quality.

**Figure 1.** Super Resolution Image Formation Model

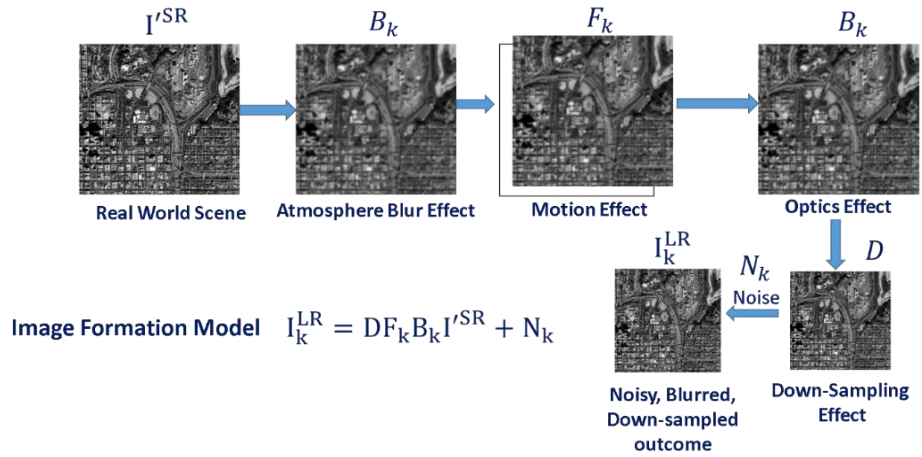

## 2 Ocean Color Monitor-3 Imaging Scheme

The Ocean Color Monitor-3 (OCM-3) utilizes an advanced imaging scheme to achieve a Signal-to-Noise Ratio (SNR) exceeding 800, employing both in-orbit and on-ground binning processes a shown in Figure 2. With a frame camera featuring 4000 across-track pixels and 48 along-track pixels, tilted at ±20 degrees to minimize sun glint, OCM-3 captures frames over 64 milliseconds, generating 24 LAC samples for each ground feature.

The large field of view (FOV) of ±43.5 degrees introduces distortions that lead to variations in Instantaneous Ground Field of View (IGFOV) across different rows, complicating data integration. As a result, the same ground object is spread across multiple pixels, with sub-pixel shifts arising from satellite motion and optical distortions a shown in Figure 3. To tackle these challenges, the on-ground OCM-3 processing employs a well-calibrated geometric model that establishes precise relationships between pixels across frames, accurately mapping image data to corresponding ground positions while accounting for motion and distortions.

Knowledge of the satellite's attitude—roll, pitch, and yaw—is essential for correcting these distortions and ensuring precise pixel registration. Although on-board binning facilitates efficient data management, it complicates pixel-to-ground correspondence. The geometric model effectively addresses these complexities, allowing OCM-3 to navigate the challenges of satellite motion and optical distortions. This results in accurate spatial representation, crucial for reliable Earth observation.

**Figure 2.** OCM-3 Imaging Scheme

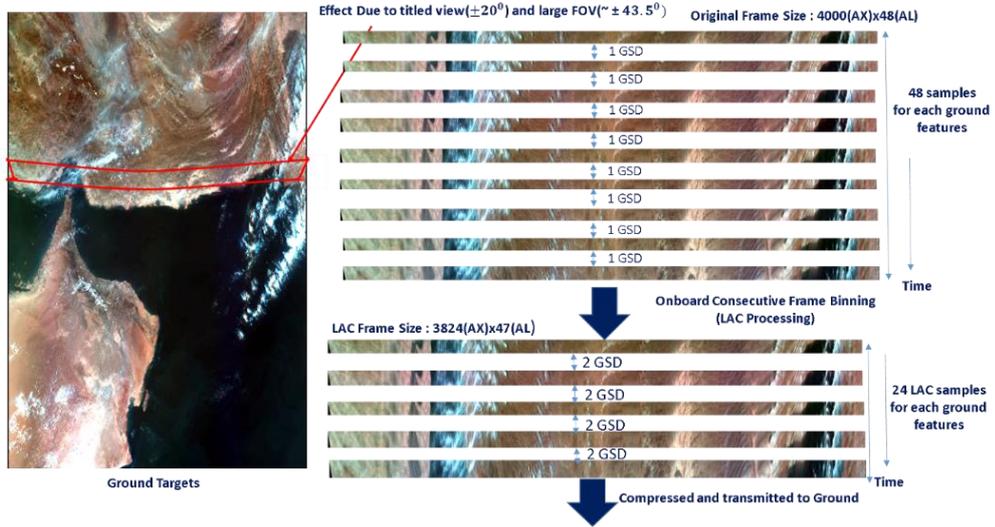

**Figure 3.** Distribution of Ground Location in the OCM-3 Frame Array

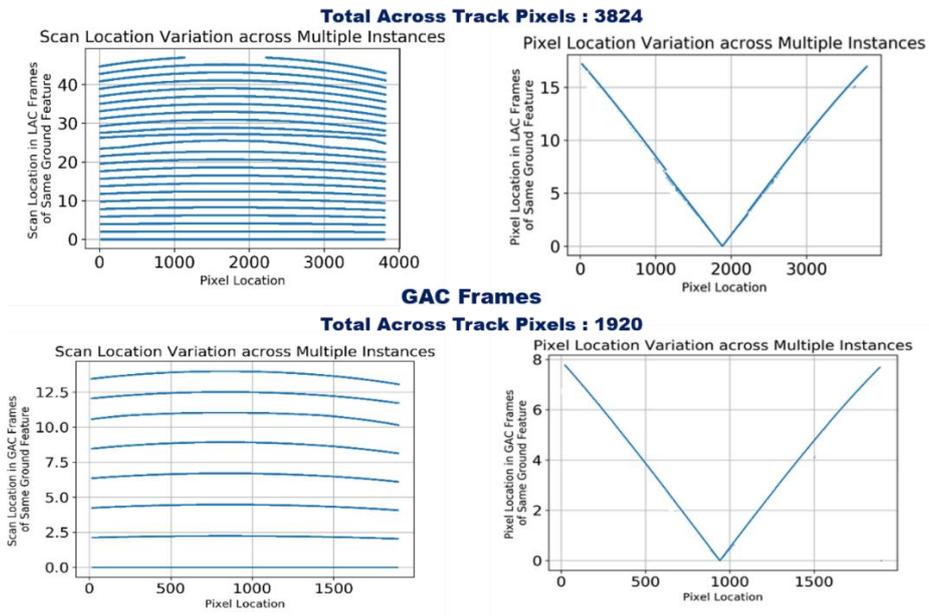

OCM-3 by its way of imaging satisfies super resolution requirement.

**Number of Samples and Phase Sampling:** OCM-3 enhances super resolution by capturing a higher number of samples than its native resolution, allowing for detailed information extraction. It incorporates phase sampling to capture sub-pixel shifts in ground feature positions across frames, which is essential for reconstructing higher-resolution images. Figure 4 illustrates these sub-pixel shifts.

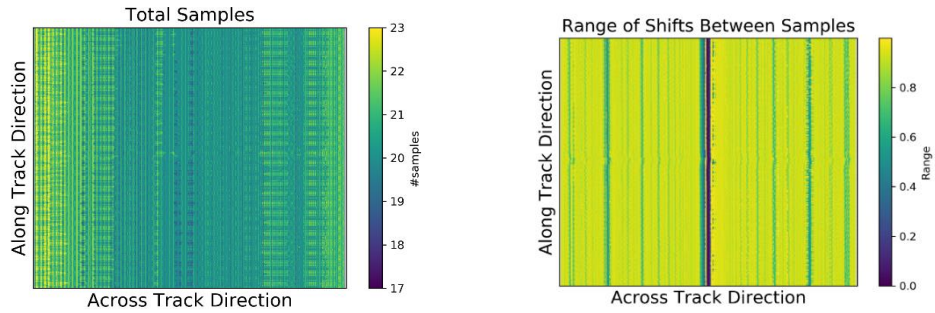

**Figure 4.** Subpixel Shifts Across Mulitple Samples in OCM-3

**Aliasing and Modulation Transfer Function (MTF):** OCM-3 mitigates aliasing, which can distort high-frequency components, by ensuring an adequate number of samples per unit length. The system maintains an MTF up to 0.75 cycles/pixel (as shown in Figure 5), enabling faithful representation of details up to this spatial frequency without significant loss of contrast. This design choice is crucial for preserving high-frequency information necessary for effective super resolution.

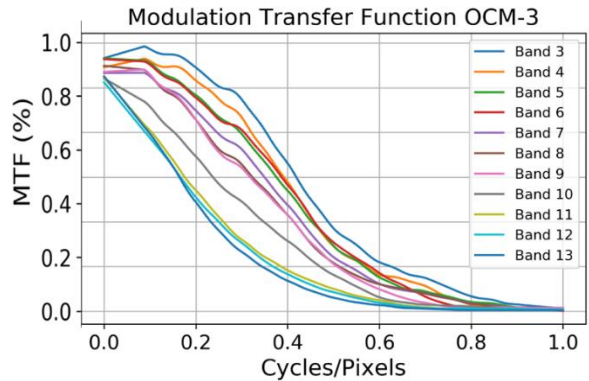

**Figure 5.** Bandwise Modulation Transfer Function of OCM-3

In summary, OCM-3 effectively addresses super resolution requirements through adequate sampling, phase sampling for sub-pixel detail, and a well-defined MTF that extends to 0.75 cycles/pixel, collectively enhancing its capability to capture and re-construct images beyond the native sensor capabilities.

## 3    Advantage of Deep Learning over Traditional Approach:

Traditional super-resolution methods utilize a straightforward approach known as "shift and fuse" In Figure 6. This involves initial motion estimation between low-resolution images to determine pixel displacement, which is crucial for aligning the images in the next step. Motion compensation ensures a coherent representation of the scene, followed by sampling onto a higher resolution grid to reconstruct a detailed version of the original scene. However, this process can lead to blurred super-resolved images due to spatial aggregation. To mitigate this, spatial aggregation techniques are employed, and subsequent image restoration steps—including deblurring and de-noising—refine the blurred representation. While effective, traditional methods have limitations, such as the need for precise motion estimation, accurate high-resolution point spread function (PSF) estimation, and handcrafted parameters for deblurring and denoising. These challenges can lead to artifacts in the final image if system knowledge is imprecise or if errors occur in estimation.

In contrast, deep learning-based super-resolution algorithms offer several advantages. They eliminate the need for handcrafted parameters, as the model learns optimal settings directly from training data, making it adaptable to varying imaging conditions. Additionally, deep learning models can perform flow estimation, capturing complex motion patterns more effectively than traditional methods. These models learn an end-to-end mapping from low-resolution to high-resolution images, streamlining the entire process without manual intervention. When trained properly, deep learning approaches can generate high-quality, artifact-free super-resolved images, which is particularly valuable for applications such as ocean-related imaging, where precision is crucial. Overall, deep learning-based methods surpass traditional hand-crafted techniques by automatically learning optimal transformations, ensuring ro-bustness and accuracy in diverse imaging scenarios.

**Figure 6.** Traditional Shift and Fuse Super Resolution Algorithm

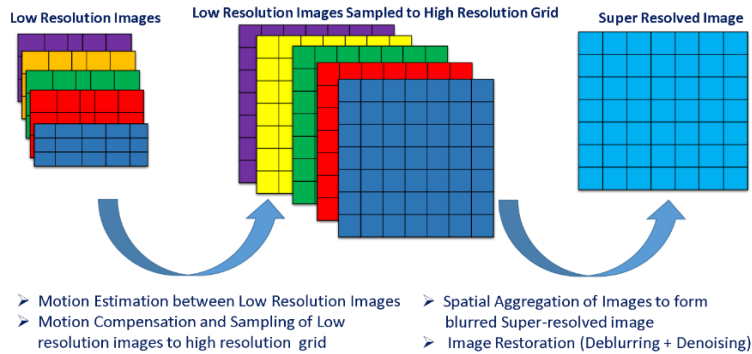

# 4 Methodology:

The developed neural network architecture is tailored for super-resolving a sequence of low-resolution images $I_T^{LR}$ to produce a single super-resolved image $I'^{SR}$. Taking inspiration from traditional shift-and-add Multiple Image Super-Resolution (MISR) algorithms similar to , the architecture encompasses three key modules: Motion Estimator, Encoder, and Decoder as shown in Figure 7. The detailed architecture of each module is shown in Figure 8.

Commencing with the Motion Estimator, motion fields between all low-resolution frames in the sequence and a reference frame $I_0^{LR}$ are estimated. Subsequently, frames undergo upscaling and alignment by compensating for motion using a Subpixel Motion Compensation (SPMC) layer. Originally designed for video SR, the SPMC layer is applied to convolutional features extracted from the frames, capitalizing on the rich local neighborhood description encoded in deep feature representations.

The upscaled and aligned features are then averaged to form a high-resolution feature map $J^{HR}$, and the super-resolved image is obtained through decoding this feature map. The network's operation can be summarized in three steps: encoding, temporal feature aggregation, and decoding. Temporal aggregation is achieved through feature averaging facilitated by a feature shift-and-add block, allowing the aggregation of an arbitrary number of frames and ensuring permutation invariance.

The trainable modules, highlighted in red, include the Motion Estimator, Encoder, and Decoder. The Motion Estimator estimates optical flows between each LR frame and the reference frame, utilizing an hourglass-style architecture. The Encoder generates relevant features for each LR image in the sequence, and the Decoder reconstructs the super-resolved image from the fused features.

A critical component is the Feature Shift-and-Add block, where features from each frame are first upscaled and motion-compensated using the SPMC module. A weighted average is then computed to map and aggregate feature pixels onto the HR grid. The SPMC module employs optical flows to compute the positions of samples in the HR grid, facilitating differentiable operations with respect to both intensities and optical flows.

The architecture's flexibility is highlighted by its permutation invariance, allowing the aggregation of an arbitrary number of frames. Notably, the Feature Shift-and-Add block does not contain any trainable parameters, contributing to its stability. In summary, the network's three-step process of encoding, temporal feature aggregation, and decoding, along with its trainable Motion Estimator, Encoder, and Decoder modules, collectively contribute to the generation of high-quality super-resolved images from a sequence of low-resolution inputs.

# 5 Training Details

The training process for the super-resolution deep learning model focuses on optimizing performance and generalization. The model is trained on approximately 15,000 64x64 pixel patches, representing pairs of high-resolution and low-resolution images, with a separate validation dataset of around 1,000 patches to assess generalization and prevent overfitting. The motion network is pre-trained on the training dataset to capture spatial and temporal transformations, aiding in motion estimation between frames. Hyper-parameter tuning is conducted via cross-validation, adjusting parameters like batch size, learning rates, and network architecture for optimal performance. The model employs L1 Loss to minimize the absolute differences between predicted and ground truth values, promoting visually appealing results. The Adam optimizer, with differentiated learning rates (1e-4 for the encoder/decoder and 1e-5 for the motion network), enhances training efficiency.

Training occurs with a batch size of 16 and spans 500 epochs using 13 V100 GPUs over 48 hours. Early stopping is implemented to prevent overfitting by halting training if validation performance does not improve after a set number of epochs. The ultimate goal is to achieve a super-resolution factor of 2, doubling the resolution of input images. Overall, this structured approach integrates data preparation, model architecture, and hyper-parameter tuning, resulting in a robust super-resolution model capable of generating high-quality images while minimizing overfitting risks.

**Figure 7.** Super Resolution Network Architecture

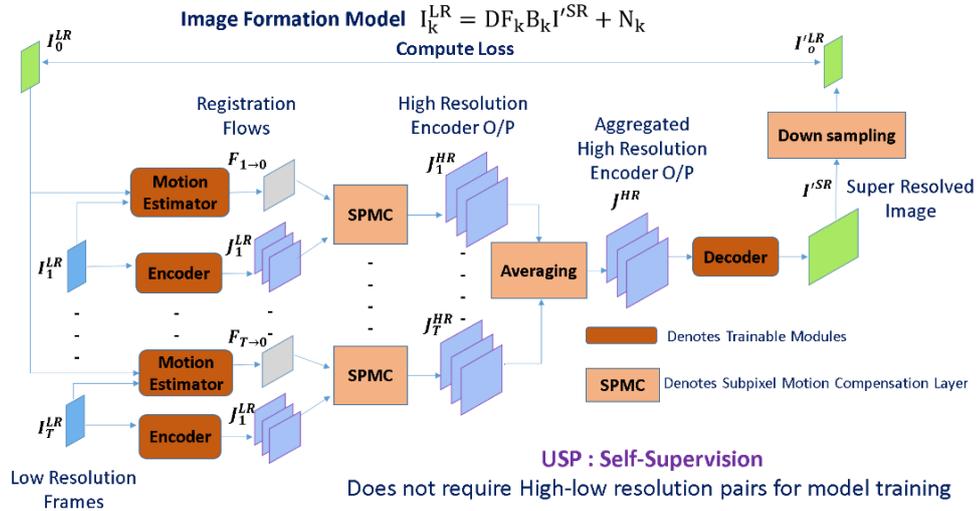

**Figure 8.** Architecture of Sub-Modules in Super Resolution Network Architecture

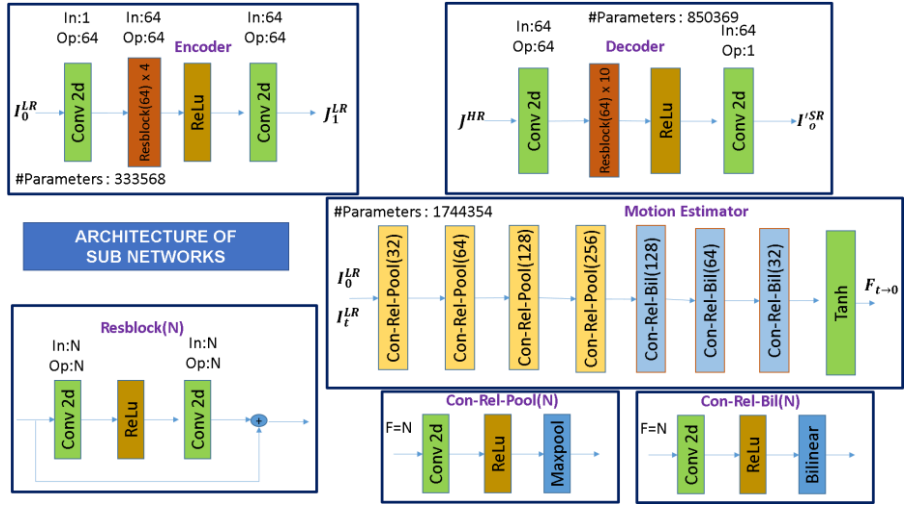

## 6      Results and Discussion

The evaluation of super-resolved images utilized a comprehensive approach, integrating various methods for thorough assessment.

### 6.1     Visual Evaluation with BRISQUE:

Visual evaluation of super-resolved images utilized the Blind/Reference less Image Spatial Quality Evaluator (BRISQUE), an objective metric for assessing perceived image quality without needing a reference. BRISQUE identifies deviations from natural image statistics, which can indicate artifacts or distortions from the super-resolution process. The BRISQUE score reflects perceived distortion: higher scores indicate lower quality and potential artifacts, while lower scores suggest better alignment with natural image statistics. This inverse relationship means that a lower BRISQUE score signifies improved image quality. Figures 9 to 12 illustrate comparisons between super-resolved images and their original resolutions, along with their BRISQUE scores. The lower BRISQUE values for the super-resolved images confirm an enhancement in visual quality post-processing.

**Figure 9.** Comparison of Low Resolution (360m) vs Super Resolved (180m) Image

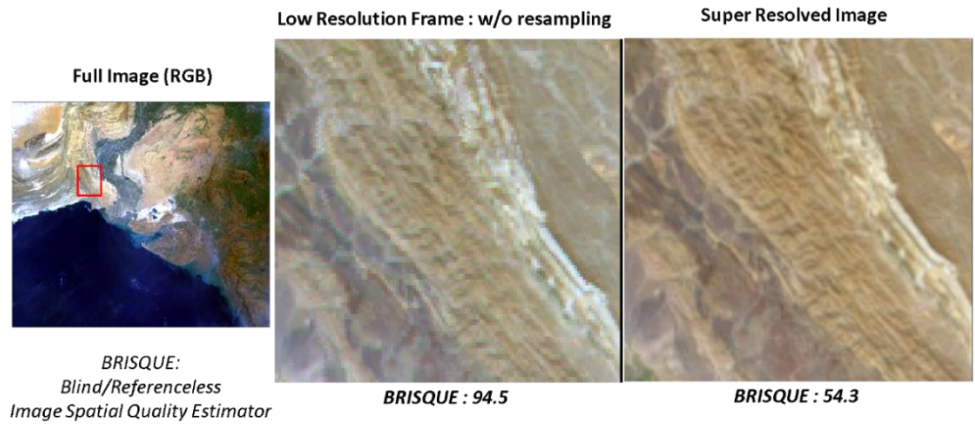

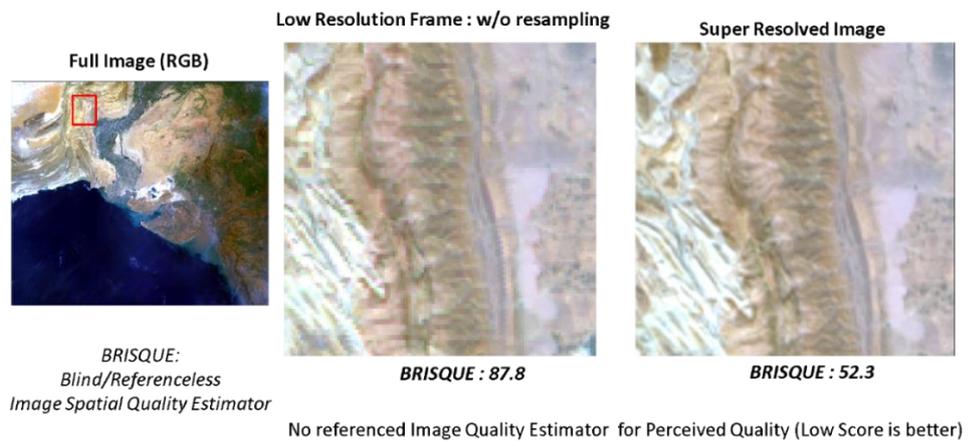

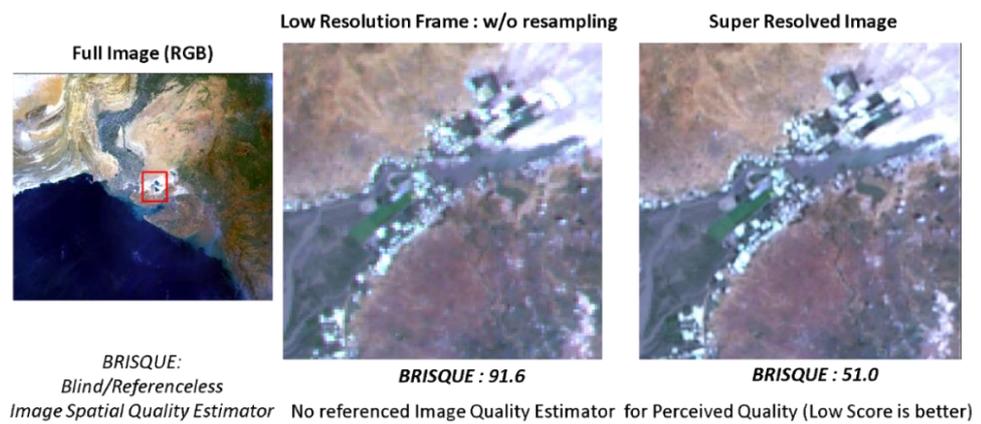

**Figure 10.** Comparison of Low Resolution (1080m) vs Super Resolved (540m) Image

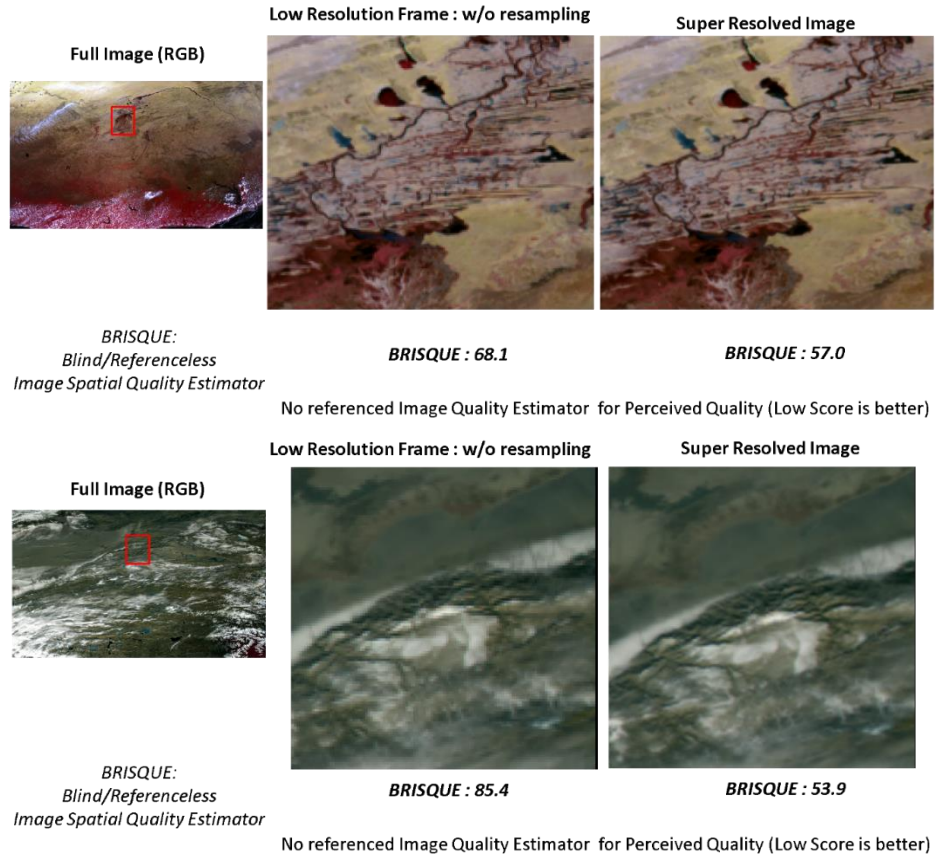

**Full Image (RGB)**

**Low Resolution Frame : w/o resampling**

**Super Resolved Image**

*BRISQUE:*
*Blind/Referenceless*
*Image Spatial Quality Estimator*

*BRISQUE : 68.1*

*BRISQUE : 57.0*

No referenced Image Quality Estimator for Perceived Quality (Low Score is better)

**Full Image (RGB)**

**Low Resolution Frame : w/o resampling**

**Super Resolved Image**

*BRISQUE:*
*Blind/Referenceless*
*Image Spatial Quality Estimator*

*BRISQUE : 85.4*

*BRISQUE : 53.9*

No referenced Image Quality Estimator for Perceived Quality (Low Score is better)

## 6.2 Super resolution factor Based on Line Spread Function

To evaluate the impact of super-resolution on image sharpness, we used Line Spread Function (LSF) and Full Width at Half Maximum (FWHM) as key metrics. Initially, FWHM was calculated for each spectral band at a ground resolution of 360 meters to establish a baseline for sharpness. The image was then interpolated to a higher resolution of 180 meters, and FWHM was recalculated. The Super Resolution (SR) Ratio, which measures the enhancement in sharpness from super-resolution compared to simple interpolation, was determined by comparing the FWHM of the super-resolved image at 180 meters with that obtained after interpolation. Results are summarized in Table 2. For this analysis, an edge over the Gujarat coast was chosen (Figure 14) due to its well-defined features, which highlight perceptible changes in sharpness. The selected edge also has a good Signal-to-Noise Ratio (SNR), ensuring reliable measurements that reflect the system's performance. In summary, the computations of LSF, FWHM, and SR Ratio provide a robust quantitative assessment of the super-resolution technique's impact on image sharpness, particularly over distinct geographic features like the Gujarat coast.

**Figure 11.** Derivation of Super Resolution Factor

| Band# | FWHM @360m | FWHM Bicubic @180m | FWHM SR @180m | SR Ratio |
|-------|-----------|--------------------|---------------|----------|
| 1 | 1.4 | 3.0 | 1.8 | 1.7 |
| 2 | 1.4 | 3.2 | 1.9 | 1.7 |
| 3 | 1.5 | 3.3 | 2.0 | 1.6 |
| 4 | 1.7 | 3.7 | 2.1 | 1.8 |
| 5 | 1.6 | 3.5 | 2.1 | 1.6 |
| 6 | 1.4 | 3.1 | 2.0 | 1.6 |
| 7 | 1.4 | 3.4 | 2.0 | 1.7 |
| 8 | 1.6 | 3.6 | 2.1 | 1.7 |
| 9 | 1.6 | 3.4 | 2.1 | 1.6 |
| 10 | 1.7 | 3.8 | 2.2 | 1.7 |
| 11 | 2.0 | 4.3 | 2.5 | 1.7 |
| 12 | 1.8 | 4.1 | 2.4 | 1.7 |
| 13 | 2.2 | 4.8 | 2.6 | 1.8 |

**Figure 12.** Line Spread Function Derivation

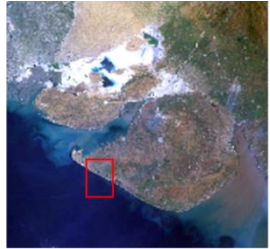

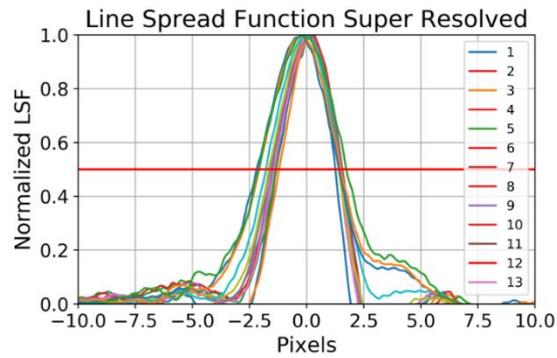

### 6.3 Statistical Comparison in terms of Power Spectrum

In addition to the LSF evaluation, we performed a statistical comparison by calculating the power spectrum of the images (Figure 15). The power spectrum is crucial for assessing spatial content, with higher power indicating more fine details and spatial frequencies. We compared the power spectrum of the super-resolved image with that of an image obtained through cubic interpolation at a resolution of 180 meters. This analysis aimed to evaluate the distribution of spatial frequencies and determine if the super-resolution process adds valuable information. Results showed a significant increase in power across nearly all spatial frequencies in the super-resolved image compared to the cubic interpolated image. This suggests that the super-resolution technique effectively captures more detailed spatial information, enhancing the image's overall quality. In summary, the power spectrum analysis provides quantitative evidence of the superior spatial content in super-resolved images. The increased power across spatial frequencies highlights the technique's ability to capture finer details, validating its effectiveness over standard interpolation methods.

**Figure 13.** Power Spectrum Comparison

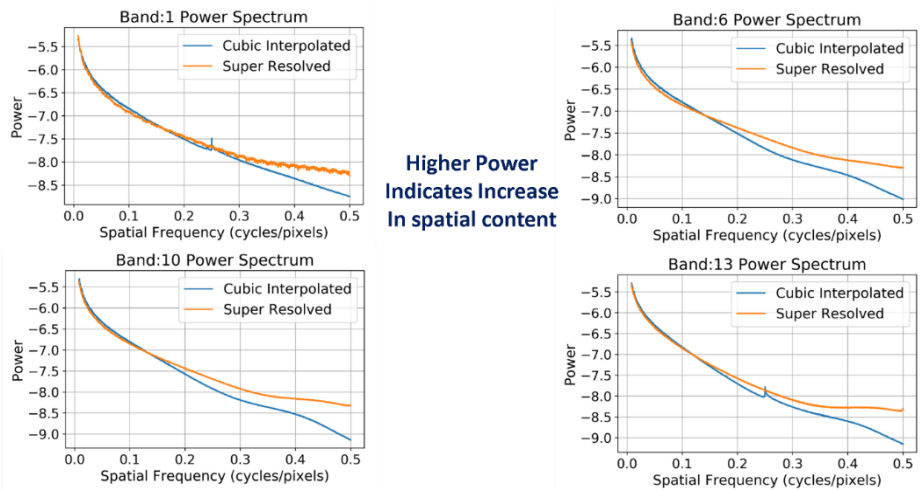

### 6.4 Targetwise Spectral Comparison

A comprehensive evaluation was conducted by comparing target-wise spectral signatures before and after the super-resolution process, focusing on areas such as deep ocean, desert, vegetation, and clouds (Figures 16 and 17). The results showed an exact match between the spectral signatures of these targets pre- and post-super-resolution. This preservation of spectral information across diverse landscapes demonstrates the robustness and accuracy of the technique in capturing the unique characteristics of various geophysical features. This finding is significant as it indicates that super-resolution does not introduce distortions to the inherent spectral properties of targets. The ability to faithfully reproduce spectral signatures enhances the reliability and applicability of the method for geospatial analyses and scientific investigations.

**Figure 14.** Spectral Comparison Before and After Super Resolution

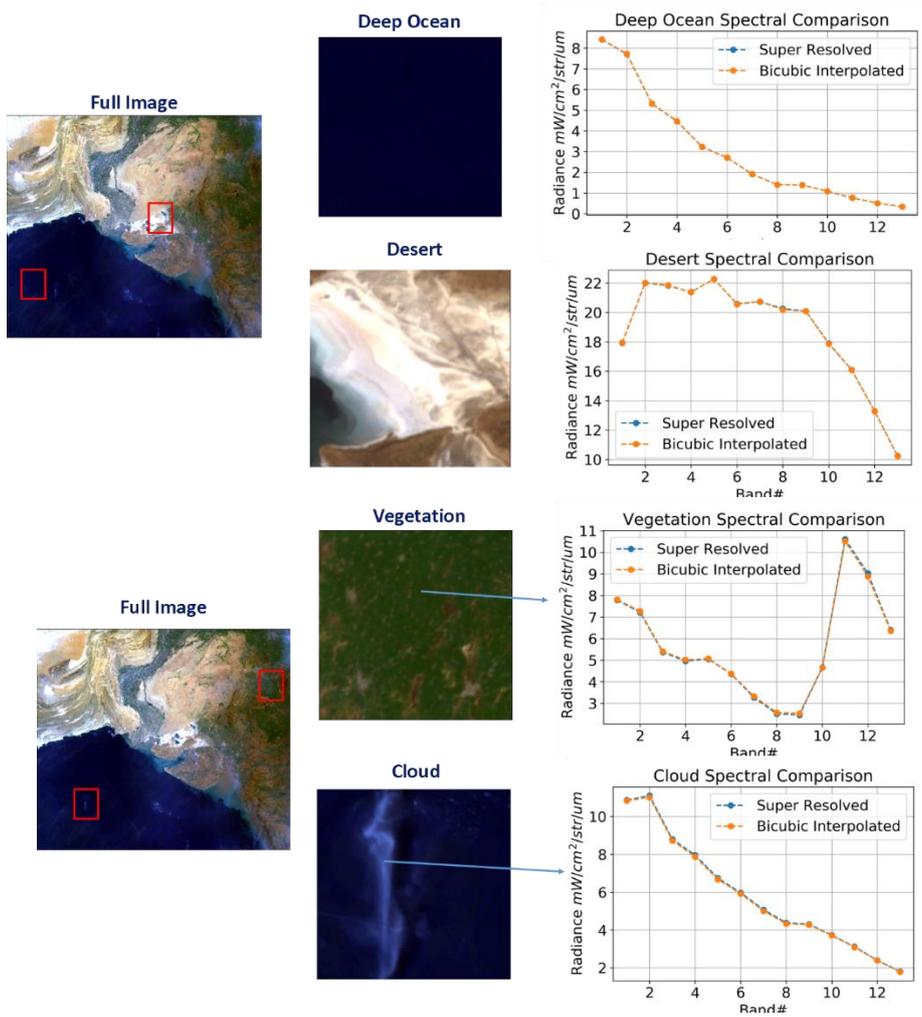

## 6.5     Evaluation of super-resolved Ocean Geophysical Parameters

The evaluation compared three types of L1B radiance datasets—single frame, super-resolution, and binned L1B datasets—at the Level 2 operational product level across six scenes from the Bay of Bengal, Arabian Sea, and North Pacific Ocean. Operational products were analyzed for chlorophyll, Diffuse Attenuation Coefficient (Kd), To-tal Suspended Matter (TSM), Aerosol Optical Depth (AOD), and Remote Sensing Reflectance (Rrs), showing a reasonable match with differences of less than 20% at most collocated points. Differences were smaller in open ocean regions (<10%) and larger in coastal areas with higher chlorophyll concentrations. The super-resolution products closely resembled operational products, while single frame products exhibited more noise and distortion. Histograms for all three datasets showed similarities, with less than 10% difference in mean values. The super-resolution product at 250m resolution effectively captured oceanic features, often sharper than the single frame images, particularly in highlighting phytoplankton blooms. Chlorophyll images from the three datasets appeared similar, but statistical analysis revealed differences. The single frame products were noisier and blurrier upon zooming, while super-resolution images were sharper. Mean and standard deviation values for chlorophyll, Kd, AOD, and TSM indicated that the super-resolution and original products had smaller differences compared to single frame products. Overall, the analysis of six OCM-3 scenes indicated that super-resolution products have accuracy comparable to the original 360m product, with more prominent oceanic features. This makes super-resolution a promising alternative for coastal and inland water applications, where improved spatial resolution enhances the retrieval of geophysical products.

**Table 2.** Operational products statistics for 11 February 2024 (Path 52 Row 14)

| Parameter | Single Frame | Super Resolved Image | L1C Image |
|---|---|---|---|
| Chlorophyll (mg/m^3) | Mean: 3.54 Std: 0.98 | Mean: 3.41 Std: 0.96 | Mean: 3.32 Std: 0.93 |
| Diffuse Attenuation coefficient (kd) m-1 | Mean: 0.35 Std: 0.22 | Mean: 0.32 Std: 0.2 | Mean: 0.31 Std: 0.2 |
| AOD | Mean: 0.18 Std: 0.1 | Mean: 0.178 Std=0.1 | Mean: 0.182 Std: 0.1 |
| TSM (mg/l) | Mean: 5.23 Std: 11.77 | Mean: 5.06 Std: 11.22 | Mean: 5.07 Std: 11.30 |

**Table 3.** Operational products statistics for 13 February (Path 64 Row 15)

| Parameter | Single Frame | Super Resolved Image | L1C Image |
|---|---|---|---|
| Chlorophyll (mg/m^3) | Mean: 0.90 Std: 0.56 | Mean: 0.88 Std: 0.42 | Mean: 0.87 Std: 0.41 |
| Diffuse Attenuation coefficient (kd) | Mean: 0.1 Std: 0.038 | Mean: 0.1 Std= 0.031 | Mean: 0.1 Std=0.03 |
| AOD | Mean: 0.29 Std=0.25 | Mean= 0.28 Std=0.26 | Mean: 0.29 Std: 0.25 |
| TSM (mg/l) | Mean: 3.40 Std: 3.98 | Mean: 3.37 Std: 3.71 | Mean: 3.38 Std: 3.75 |

**Table 4.** Operational products statistics for 13 February(Path 64 Row 15)

| Parameter | Single Frame | Super Resolved Image | L1C Image |
|---|---|---|---|
| Chlorophyll (mg/m^3) | Mean: 0.89 Std: 0.58 | Mean: 0.85 Std: 0.42 | Mean: 0.85 Std: 0.42 |
| Diffuse Attenuation coefficient (kd) | Mean : 0.1 Std: 0.039 | Mean : 0.098 Std: 0.032 | Mean: 0.098 Std: 0.032 |
| AOD | Mean: 0.27 Std: 0.28 | Mean: 0.27 Std: 0.28 | Mean: 0.27 Std: 0.28 |
| TSM (mg/l) | Mean: 3.26 Std: 4.61 | Mean: 3.23 Std: 4.38 | Mean: 3.24 Std: 4.48 |

**Figure 15.** Chlorophyll images

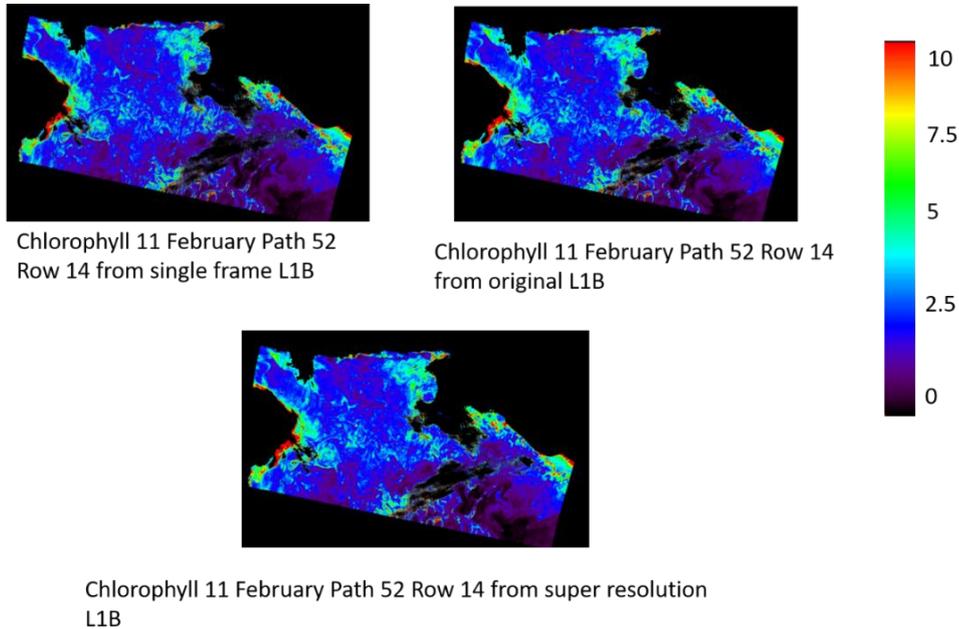

Chlorophyll 11 February Path 52
Row 14 from single frame L1B

Chlorophyll 11 February Path 52 Row 14
from original L1B

Chlorophyll 11 February Path 52 Row 14 from super resolution
L1B

### 6.6 Evaluation of Super-resolved Cryosphere Geophysical Parameters

The cryosphere, encompassing sea ice, glaciers, snow cover, and permafrost, plays a vital role in regulating Earth's climate system. Monitoring changes in these dynamic components requires high-resolution satellite data to accurately capture crucial details. However, acquiring high-resolution imagery frequently is impractical due to limitations in sensor capabilities or revisit times. Super-resolution techniques offer a promising alternative by enhancing the spatial resolution of existing low-resolution images, potentially unlocking new insights into cryosphere processes.

This report presents an evaluation of OCM-3 TOA reflectance of 262 m super-resolution data generated from OCM-3 TOA radiance LAC (Local Area Coverage) of 360 m resolution data acquired on 28th November 2023. We have evaluated the effectiveness of the super resolution data with reference of OCM LAC data. As pre-processing part, we have applied a radiometric correction on both images and compared reflectance values on different features as shown in Figure 21. Comparison of TOA reflectance for Bands 5, 8 and 13 of both the datasets are given in Table 6. There is a significant, strong and positive correlation of 0.99 between TOA Reflectance of OCM-3 super resolution data and TOA reflectance of OCM-3 LAC data Bands 5, 8 & 13 .Some features like rift, crevasse highlight more efficiently in super resolution data compared with LAC data as shown in Figure 22. TOA reflectance values of ocean and melt are nearly same for both the dataset in for selected bands while on the ice features like ice-sheet, ice-shelf and sea-ice, there are minor deviations.

**Figure 16.** False Colour Composite image generated using B5(550-560nm), B8(665-675nm) and B13(990-1030nm) of OCM-3 TOA reflectance super resolution data highlighting the region of interest of different features for comparison of reflectance values with OCM-3 TOA

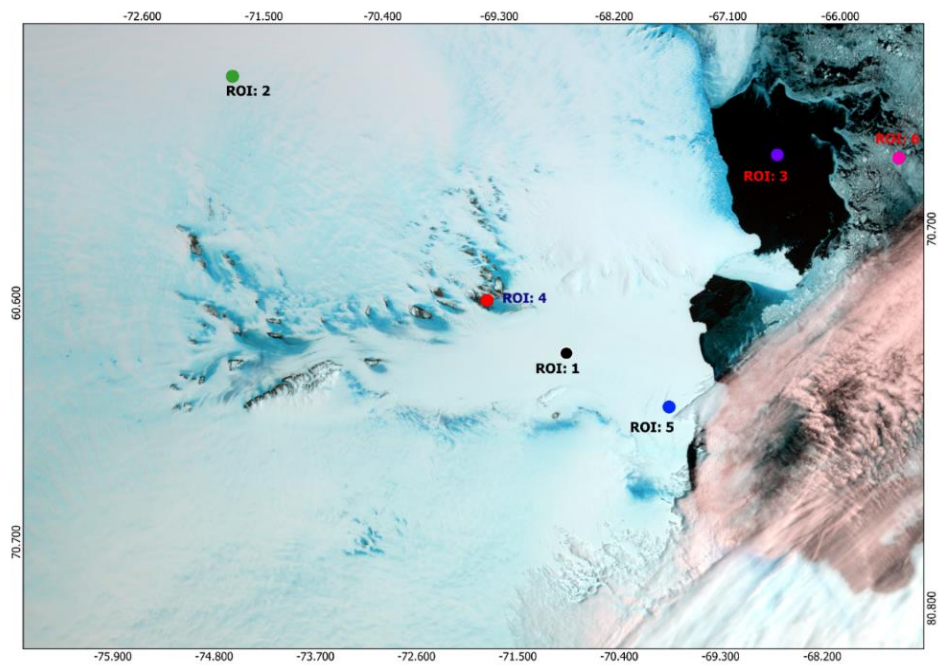

**Figure 17.** Scatterplot by band of OCM-3 super resolution TOA reflectance data versus OCM-3 LAC TOA reflectance data

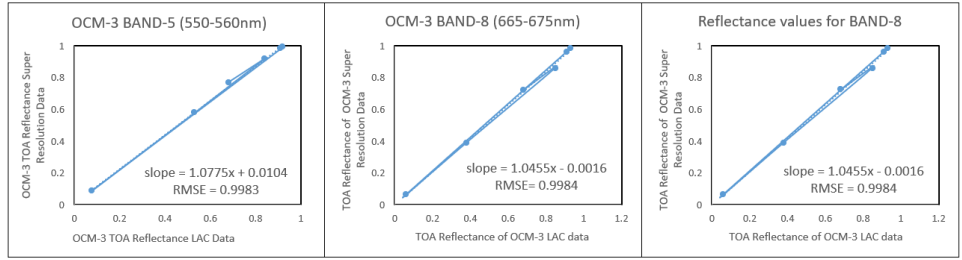

**Figure 18.** FCC Image generated using B5, B8 & B13 of OCM-3 TOA reflectance super resolution Data, B:FCC Image generated using B5, B8 & B13 of OCM-3 TOA reflectance LAC Data, highlighting a rifts over Amery Ice-Shelf, Antarctic

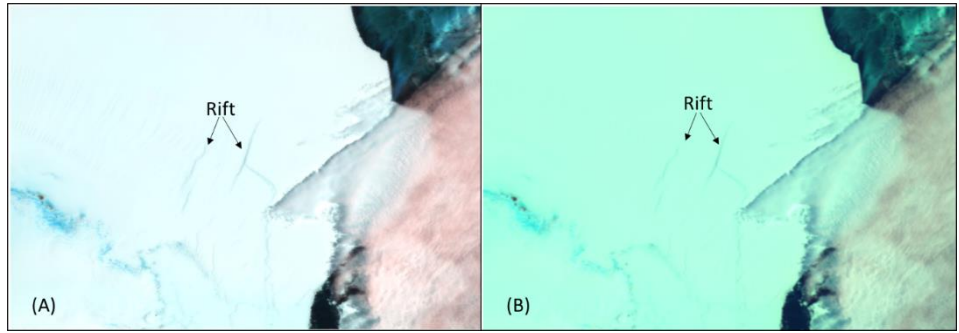

**Table 5.** Comparison between TOA Reflectance of OCM-3 super resolution data and OCM-3 LAC data over different region of interest as shown in Figure :20

| S. No. | Features | TOA Reflectance SOCM-3 | | | TOA Reflectance OCM-3 LAC | | |
|--------|----------|------|------|------|------|------|------|
| | **ROI** | **B5** | **B8** | **B13** | **B5** | **B8** | **B13** |
| 1 | Ice-Shelf (ROI-1) | 0.91 | 0.93 | 0.74 | 0.98 | 0.98 | 0.69 |
| 2 | Ice-Shelf (ROI-2) | 0.92 | 0.91 | 0.69 | 0.99 | 0.96 | 0.65 |
| 3 | Ocean (ROI:3) | 0.076 | 0.060 | 0.053 | 0.085 | 0.064 | 0.05 |
| 4 | Melt (ROI:4) | 0.53 | 0.38 | 0.07 | 0.58 | 0.39 | 0.06 |
| 5 | Rift (ROI:5) | 0.84 | 0.85 | 0.62 | 0.92 | 0.86 | 0.56 |
| 6 | Sea Ice (ROI:6) | 0.68 | 0.68 | 0.45 | 0.77 | 0.72 | 0.44 |

### 6.7    Evaluation of Super-resolved Image Over Vegetation Target

An in-depth analysis was conducted on sample OCM-03 datasets, comparing super-resolved images with their original counterparts. Additionally, a comprehensive comparative study focused on NDVI datasets was carried out, specifically targeting the dominant crop areas of Madhya Pradesh. The results of the analysis revealed intriguing insights. Transects drawn from these datasets indicated that NDVI computed from super-resolved images effectively captured the NDVI pattern when compared to the original NDVI images. However, a nuanced observation emerged as super-resolved NDVI datasets exhibited higher sensitivity at specific NDVI values than their original counterparts. This heightened sensitivity suggests the potential inclusion of additional information in the super-resolution datasets. The transects depicted in the NDVI images, comparing super-resolution and original resolution datasets, are visually presented in the following graph. This comparative analysis highlights the enhanced sensitivity and potential wealth of information offered by super-resolution datasets, providing a valuable perspective for applications in agricultural monitoring and environmental studies.

**Figure 19.** NDVI Profile of Super Resolved Images

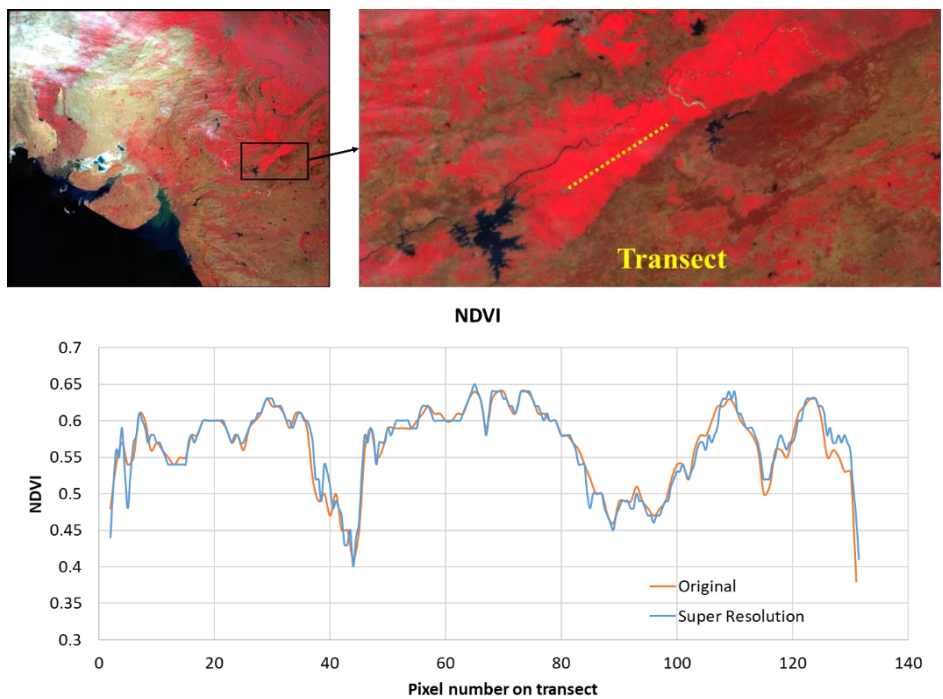

## Conclusion

This study highlights the effectiveness of super-resolution techniques in enhancing the quality of EOS-06 OCM-3 data. Through comprehensive evaluations—including visual assessments, computational metrics, power spectrum analysis, and spectral comparisons—we demonstrated that super-resolution algorithms significantly improve image sharpness and detail retention.

The analysis revealed that super-resolved images exhibited superior spatial content, with a marked increase in clarity, particularly in coastal and inland water applications. Notably, the spectral signatures of various geophysical features remained consistent post-super-resolution, indicating that the algorithms do not introduce distortions.

The findings suggest that super-resolution not only enhances the visual quality of imagery but also preserves the integrity of the spectral information essential for accurate remote sensing analyses. This advancement presents a valuable tool for scientists and researchers, facilitating more precise observations and interpretations in environmental monitoring and resource management. Overall, the application of super-resolution algorithms offers a promising pathway for improving the utility of satellite data in various scientific and practical contexts.

## Acknowledgment

The authors thankfully acknowledges the understanding, encouragement and support received from Director, Space Applications Centre, ISRO. The authors would like to thank Shri. Debajyoti Dhar, Deputy Director, Signal & Image Processing Area and Mrs. Rashmi Sarma, Deputy Director, Earth and Planetary Sciences and Application Area. The authors would also like to thank other SIPA and EPSA members who have given their support from time to time. The continuing support from IRS project Management through IRS Program Director, Oceansat-3 Project Director, Associate Project Director, Payloads and Ground Segment Committee members is thankfully acknowledged.

## Funding N/A

## Declarations

**Conflict of Interest** The authors declared that they have no conflict of interest.

# References

1. Irani, M., Peleg, S.: Improving resolution by image registration. CVGIP: Graphical models and image processing 53(3), 231–239 (1991).
2. Stark, H., Oskoui, P.: High resolution image recovery from image-plane arrays, using convex projections. J. Opt. Soc. Am. A 6(1989), 1715–1726.
3. Lee, E., Kang, M.: Regularized adaptive high-resolution image reconstruction considering inaccurate subpixel registration. IEEE Trans. Image Process. 12(2003), 806–813.
4. Ledig, C., Theis, L., Huszár, F., Caballero, J., Cunningham, A., Acosta, A., Aitken, A., Tejani, A., Totz, J., Wang, Z., Shi, W.: Photo-realistic single image super-resolution using a generative adversarial network. In: CVPR 2017, pp. 10.1109/CVPR.2017.19 (2017).
5. Farsiu, S., Robinson, M.D., Elad, M., Milanfar, P.: Fast and robust multiframe super-resolution. IEEE Trans. Image Process. 13(10), 1327–1344 (2004).
6. Schultz, R., Stevenson, R.: Extraction of high-resolution frames from video sequences. IEEE Trans. Image Process. 5(6), 996–1011 (1996).
7. Ng, M., Shen, H., Lam, E., Zhang, L.: A total variation regularization based super-resolution reconstruction algorithm for digital video. J. Adv. Signal Process. (2007), 1–16.
8. Zhang, X., et al.: Application of Tikhonov regularization to super-resolution reconstruction of brain MRI images. In: Medical Imaging and Informatics 2007, pp. 1–8. Springer, Berlin, Heidelberg (2007).
9. Yuan, Q., Zhang, L., Shen, H.: Multiframe super-resolution employing a spatially weighted total variation model. IEEE Trans. Circuits Syst. Video Technol. 22(3), 379–392 (2011).
10. Liu, X., Zhao, J.: Robust multi-frame super-resolution with adaptive norm choice and difference curvature based BTV regularization. In: 2017 IEEE Global Conference on Signal and Information Processing (GlobalSIP), pp. 1–5. IEEE, 2017.
11. N. L. Nguyen, J. Anger, A. Davy, P. Arias and G. Facciolo, "Self-supervised multi-image super-resolution for push-frame satellite images," 2021 IEEE/CVF Conference on Computer Vision and Pattern Recognition Workshops (CVPRW), Nashville, TN, USA, 2021, pp. 1121-1131, doi: 10.1109/CVPRW53098.2021.00123.
12. Tao, Xin, et al. "Detail-revealing deep video super-resolution." Proceedings of the IEEE international conference on computer vision. 2017.
13. Salvetti, Francesco, et al. "Multi-image super resolution of remotely sensed images using residual attention deep neural networks." Remote Sensing 12.14 (2020): 2207.
14. Salgueiro Romero, Luis, Javier Marcello, and Verónica Vilaplana. "Super-resolution of sentinel-2 imagery using generative adversarial networks." Remote Sensing 12.15 (2020): 2424.